\documentclass{iau}

\usepackage{amsmath}
\usepackage{graphicx}
\usepackage{multirow}
\usepackage{orcidlink} % added by Simone
\usepackage{subfigure}  % added by Simone

\begin{document}

\lefttitle{S. Riggi et al.}
\righttitle{Proceedings IAU Symposium}

\jnlPage{1}{4}
\jnlDoiYr{2025}
\doival{10.1017/S1743921325105085}
\volno{397}
\pubYr{2025}

\aopheadtitle{Proceedings IAU Symposium}
%\editors{XXX, YYY \& ZZZ, eds.}

\title{Toward Vision-Language Assistants for Radio Astronomical Source Analysis}

\author{S. Riggi \orcidlink{0000-0001-6368-8330}}
\affiliation{INAF - Osservatorio Astrofisico di Catania, Via Santa Sofia 78, 95123 Catania, Italy \\ email: {\tt \email{simone.riggi@inaf.it}}}

%\author{T. Cecconello}
%\affiliation{INAF - Osservatorio Astrofisico di Catania, Via Santa Sofia 78, 95123 Catania, Italy}
%\affiliation{Department of Electrical, Electronic and Computer Engineering, University of Catania, Santa Sofia 64, 95123, Catania, Italy}

\begin{abstract}
Vision-language models (VLMs) have recently shown promise in general-purpose reasoning tasks, yet their applicability to domain-specific scientific workflows remains largely unexplored. In this work, we evaluated a series of open-weight and commercial VLMs on six tasks relevant to radio astronomy, such as source morphology classification. We also introduced \texttt{radio-llava}, a fine-tuned multimodal assistant built on the LLaVA architecture and adapted for the radio domain through instruction fine-tuning. 

In zero-shot mode, commercial models like GPT-4.1 outperform open-weight VLMs on most radio benchmarks. However, \texttt{radio-llava} significantly improves upon both base LLaVA and commercial models across nearly all tasks. Despite these gains, specialized vision-only models still deliver substantially better performance across the board. Additionally, we observed that fine-tuning introduces catastrophic forgetting on general multimodal tasks, with performance drops up to 40\% that can be partly mitigated with a more diverse training dataset or shallow fine-tuning.

%Vision-language models (VLMs) have recently shown promise in general-purpose reasoning tasks, yet their applicability to domain-specific scientific workflows remains largely unexplored. In this work, we evaluated a series of open-weight and commercial VLMs — including LLaVA models of varying sizes — on six tasks relevant to radio astronomy, such as source morphology classification. We then introduced \texttt{radio-llava}, a fine-tuned multimodal assistant built on the LLaVA architecture and adapted for the radio domain. Fine-tuned models were evaluated on radio benchmarks using different training datasets and strategies, and compared with both base and vision-only models. 

%In zero-shot mode, commercial models like GPT-4.1 outperform open-weight VLMs on most radio benchmarks. However, \texttt{radio-llava} significantly improves upon both base LLaVA and commercial models across nearly all tasks. Despite these gains, specialized vision-only models still deliver substantially better performance across the board. Additionally, we observed that fine-tuning introduces catastrophic forgetting on general multimodal tasks, with performance drops up to 40\% that can be partly mitigated with a more diverse training dataset or shallow fine-tuning.
\end{abstract}


\maketitle

\section{Introduction}
Vision-language models (VLMs) like \texttt{CLIP} \citep{CLIP} and \texttt{LLaVA} \citep{LLaVA} have opened new directions in cross-modal learning by enabling natural language understanding grounded in visual inputs. Astronomy, with its rich collection of images, spectra, and tabular data, is an ideal domain for testing such capabilities. While \texttt{CLIP}-based works \citep{Gupta2025} primarily focused on static alignment for image-text retrieval tasks, this paper explores the application of generative, instruction-following VLMs to radio astronomical analysis.

\section{Methodology}
We used labelled radio image data from various surveys $-$ including MeerKAT SMGPS \citep{Goedhart2024}, ASKAP EMU \citep{EMUMainSurvey,Norris2021}, and VLA FIRST \citep{Becker1995} $-$ to construct six radio classification tasks for VLM evaluation, summarised in Table~\ref{tab:radio-benchmarks}.

\begin{table}[htb]
\centering%
%\scriptsize%
\footnotesize%
\caption{Radio benchmarks.}
\begin{tabular}{llllll}
\hline%
\hline%
ID & Task Description & Survey & \#Images & Classification & Labels\\%
\hline%
%\texttt{B1} & extended/diffuse source detection & SMGPS/ASKAP-EMU & 5,718 & multi-label & {\tiny{\texttt{EXTENDED}, \texttt{DIFFUSE}, \texttt{DIFFUSE-LARGE}}}\\%
\texttt{B1} & extended/diffuse & \texttt{SMGPS}, & 5,718 & multi-label & {\scriptsize{\texttt{EXTENDED}, \texttt{DIFFUSE}}}\\%
 & source detection & \texttt{EMU} & & & {\scriptsize{\texttt{DIFFUSE-LARGE}}}\\%
\hline%
%\texttt{B2} & source morphology classification & VLA-FIRST & 3,835 & single-label & {\tiny{\texttt{1C-1P}, \texttt{1C-2P}, \texttt{1C-3P}, \texttt{2C-2P}, \texttt{2C-3P},\texttt{3C-3P}}}\\%
\texttt{B2} & source morphology  & \texttt{FIRST} & 3,835 & multi-class & {\scriptsize{\texttt{1C-1P}, \texttt{1C-2P}, \texttt{1C-3P}, }}\\%
 & classification &  &  & single-label & {\scriptsize{\texttt{2C-2P}, \texttt{2C-3P}, \texttt{3C-3P}}}\\%
\hline%
%\texttt{B3} & extended radio-galaxy detection & SMGPS/ASKAP-EMU & 5,718 & binary & {\tiny{\texttt{YES}, \texttt{NO}}}\\%
\texttt{B3} & extended radio-galaxy & \texttt{SMGPS}, & 5,718 & binary class & {\scriptsize{\texttt{YES}, \texttt{NO}}}\\%
 & detection & \texttt{EMU} & & & \\%
\hline%
%\texttt{B4} & imaging artefact detection & SMGPS/ASKAP-EMU & 5,718 & binary & {\tiny{\texttt{YES}, \texttt{NO}}}\\%
\texttt{B4} & imaging artefact & \texttt{SMGPS}, & 5,718 & binary class & {\scriptsize{\texttt{YES}, \texttt{NO}}}\\%
& detection & \texttt{EMU} &  &  & \\%
\hline%
%\texttt{B5} & image peculiarity classification  & SMGPS/ASKAP-EMU & 5,718 & single-label & {\tiny{\texttt{ORDINARY}, \texttt{COMPLEX}, \texttt{PECULIAR}}}\\%
\texttt{B5} & image peculiarity & \texttt{SMGPS}, & 5,718 & multi-class & {\scriptsize{\texttt{ORDINARY}, \texttt{COMPLEX}, }}\\%
 & classification  & \texttt{EMU} &  &  & {\scriptsize{\texttt{PECULIAR}}}\\%
\hline%
%\texttt{B6} & radio-galaxy morphology classification & VLA-FIRST & 833 & binary & {\tiny{\texttt{FR-I}, \texttt{FR-II}}} \\%
\texttt{B6} & radio-galaxy morphology & \texttt{FIRST} & 833 & binary class & {\scriptsize{\texttt{FR-I}, \texttt{FR-II}}} \\%
& classification &  &  &  &  \\%
\hline%
\hline%
\end{tabular}
\label{tab:radio-benchmarks}
\end{table}

We first assessed the zero-shot performance of several open-weight models $-$ \texttt{LLaVA-OneVision} \citep{LLaVA-OneVision}, \texttt{Qwen2VL} \citep{Qwen2VL}, \texttt{InternVL} \citep{InternVL}, \texttt{TinyLLaVA} \citep{TinyLLaVA} $-$ of different sizes (0.5B to 72B parameters) and a representative commercial closed-weight solution (OpenAI GPT-4.1). 

We then evaluated a fine-tuned \texttt{LLaVA-OneVision} 7B model $-$ hereafter referred to as \textit{radio-llava} $-$ trained on curated Q\&A instructions derived from an independent set of 59k annotated radio images, as well as on curated 39k image-caption pairs extracted from a corpus of published radio astronomy papers (2000–2025). In all training runs we kept the vision encoder (\textit{siglip-so400m-patch14-384}) frozen, while we fully fine-tuned the LLM (\textit{qwen2}) and adapter (\textit{mlp2x\_gelu}) components.
Fine-tuning results were compared with both the base \texttt{LLaVA} model and a vision-only baseline.

Finally, we conducted additional diagnostic analyses to explore alternative \texttt{LLaVA} hyperparameter settings and to evaluate the extent of catastrophic forgetting of previously learned multimodal tasks following fine-tuning.

Further details on the training/evaluation data and strategy can be found in \cite{Riggi2025}.

\begin{figure*}[!htb]
\centering%
\includegraphics[scale=0.7]{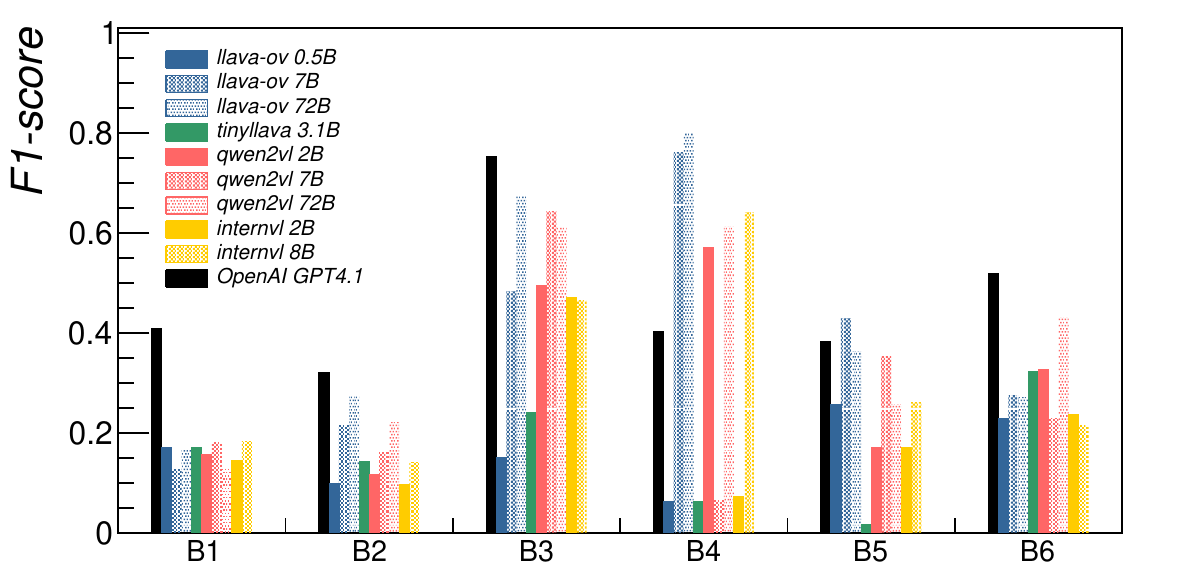}
\vspace{-0.4cm}%
\caption{
Classification "macro-averaged" F1-score obtained across tasks B1–B6 in zero-shot mode with 
%Zero-shot benchmark results (classification macro F1-score across tasks B1–B6) for 
open-weight VLMs of different sizes (0.5B, 2B, 3.1B, 7B, 8B, 72B), shown with coloured histograms (LLaVA: blue, TinyLLaVA: green, Qwen2VL: red, InternVL: orange), and OpenAI GPT-4.1 (black histograms).}%
\label{fig:eval-zeroshot}
\end{figure*}

\section{Results}
Zero-shot benchmark results are presented in Fig.~\ref{fig:eval-zeroshot}. GPT-4.1 (black histograms) outperforms the open-weight models in four out of six benchmarks, with a $\sim$20\% improvement in tasks B1 and B6. However, it unexpectedly underperforms in tasks B4 and B5. Among the open-weight models, the largest versions of \texttt{LLaVA} match or slightly exceed the performance of their counterparts. Overall, all models demonstrate relatively low performance, indicating the need for domain-specific specialization on radio astronomical data.

\begin{figure*}[!htb]
\centering%
\includegraphics[scale=0.7]{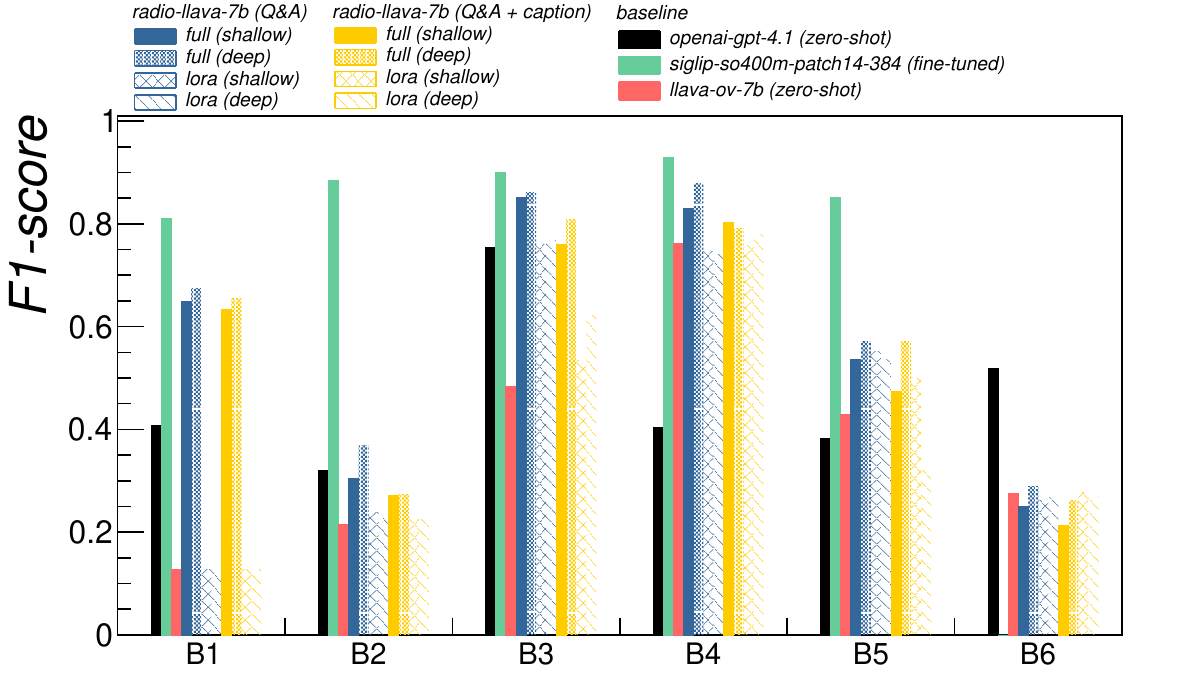}
\vspace{-0.4cm}%
\caption{
Classification "macro-averaged" F1-score obtained across tasks B1–B6 with the \textit{radio-llava}
%Benchmark results (classification F1-score across tasks B1–B6) for the \textit{radio-llava}
model, fine-tuned on the Q\&A training dataset (blue histograms) and the combined Q\&A and caption datasets (orange histograms) with different training strategies (full vs. LoRA fine-tuning) and depths (shallow vs. deep). Results from different baseline models are also shown: base \texttt{llava-ov-7b} (red histograms), fine-tuned \texttt{siglip-so400m-patch14-384} vision encoder (green histograms), OpenAI GPT-4.1 (black histograms).}%
\label{fig:eval-finetune}
\end{figure*}

Results for the fine-tuned \texttt{radio-llava} models, reported in Fig.~\ref{fig:eval-finetune} as blue and orange histograms for Q\&A-only and combined training datasets respectively, indicate substantial improvements over the base VLMs (red histograms), with gains of $\sim$30\% in tasks B1 and B3. For the remaining tasks, the improvements are more modest $-$ around 10\% in B2, B4, and B5 $-$ and negligible in B6. With the exception of B6, the fine-tuned models also outperform GPT-4.1 (black histograms), though with smaller margins. Notably, VLMs trained jointly on all tasks consistently underperform relative to vision-only classifiers trained individually per task, as shown by the green histograms in Fig.~\ref{fig:eval-finetune}.

Models fine-tuned using alternative LLaVA hyperparameters $-$ such as learning rate and relative scheduling, batch size, and LoRA configuration $-$ exhibited performance broadly comparable to that obtained with the default settings. A modest improvement ($\sim$2\%) was observed when the vision encoder was also fully fine-tuned. 

\begin{figure*}[!htb]
\centering%
\includegraphics[scale=0.7]{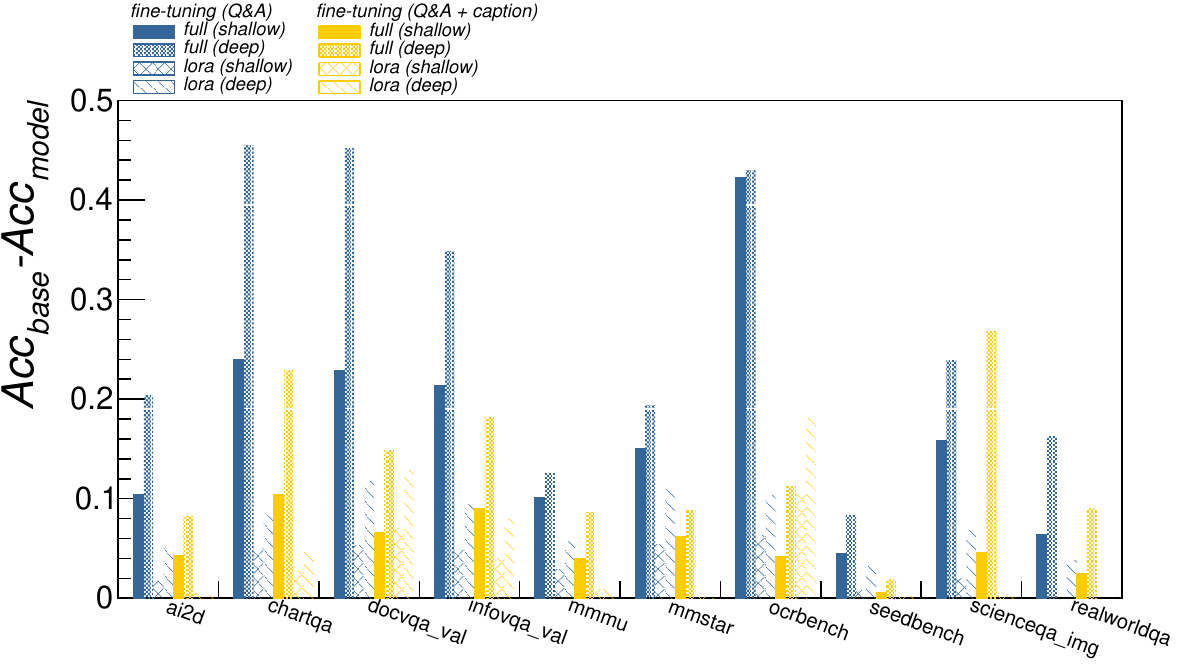}
\vspace{-0.4cm}%
\caption{
Classification accuracy differences between the \texttt{llava-ov-7b} base model and the fine-tuned \texttt{radio-llava} models (blue and orange histograms), evaluated on various multimodal (non-radio) benchmarks.}%
\label{fig:eval-multimodal}
\end{figure*}

Deeper fine-tuning (e.g., three epochs vs a single epoch) had only a minor impact (few percent) on classification performance for most radio benchmarks. However, it led to substantial degradation in performance on standard multimodal tasks $-$ such as \textit{ChartQA}, \textit{DocVQA}, and \textit{InfoVQA} $-$ with accuracy drops of up to 40\%, indicative of catastrophic forgetting, as shown in Fig.\ref{fig:eval-multimodal}. Incorporating more diverse training data, such as curated caption datasets, or adopting LoRA-based fine-tuning strategies partially mitigated this effect and helped recover some of the lost general multimodal capabilities.

\section{Summary}
%Our experiments demonstrate that while large commercial models offer impressive out-of-the-box performance, open-weight models like \texttt{radio-llava} can be fine-tuned to achieve strong results in specialized tasks and enable assistant-style interaction. Future work includes expanding datasets, improving captioning capabilities, and integrating real-time assistant interfaces for astronomers.

This work investigates the feasibility of using small-scale Vision-Language Models (VLMs) as domain-specific assistants for radio astronomical image analysis. Our evaluation shows that large commercial models like GPT-4.1 deliver strong zero-shot results on some tasks but consistently underperform compared to specialized vision-only models trained specifically for radio astronomy.

Fine-tuning smaller open-weight models like \texttt{LLaVA} on curated radio datasets (Q\&A and captions) significantly improves performance over both base and commercial models, with up to $\sim$30\% F1-score gains in certain benchmarks. However, these fine-tuned models still lag behind vision-only models trained per task and suffer from catastrophic forgetting, losing up to 40\% accuracy on general multimodal benchmarks after full deeper fine-tuning. Incorporating caption data and adopting LoRA-based fine-tuning helps mitigate some of these issues, improving instruction-following ability and partially recovering lost generalization.

%We find that the main limitations in VLM performance stem from insufficient training data size, limited diversity and label quality, and suboptimal vision-language alignment. Incorporating caption data and adopting LoRA-based fine-tuning helps mitigate some of these issues, improving instruction-following ability and partially recovering lost generalization.

%These results underscore the potential role of specialized VLMs in supporting astronomers through assistant-style interaction, and highlight the need for better multimodal alignment, more diverse training data, and improved fine-tuning strategies to fully realize their value in scientific workflows.

These results underscore both the potential and current limitations of VLMs in supporting astronomers through assistant-style interaction, and emphasize the need for better multimodal alignment, more diverse and high-quality training data, and improved fine-tuning strategies to fully realize their value in scientific workflows.

\begin{thebibliography}{200}
\bibitem[\protect\citeauthoryear{Becker et al.}{1995}]{Becker1995}%
Becker, R.~H., et al., 1995, ApJ, 450, 559, \url{https://doi.org/10.1086/176166}%
\bibitem[\protect\citeauthoryear{Chen et al.}{2024}]{InternVL}%
Chen, Z., et al., Proc. of the IEEE/CVF Conference on Computer Vision and Pattern Recognition, 24185--24198, 2024, \url{https://doi.org/10.48550/arXiv.2312.14238}
\bibitem[\protect\citeauthoryear{Goedhart et al.}{2024}]{Goedhart2024} 
Goedhart, S., et al., 2024, MNRAS, 531, 649, \url{https://doi.org/10.1093/mnras/stae1166}%
\bibitem[\protect\citeauthoryear{Gupta et al}{2025}]{Gupta2025}
Gupta, N., et al., 2025, PASA, in press, \url{https://doi.org/10.1017/pasa.2025.10064}%
\bibitem[\protect\citeauthoryear{Hopkins et al.}{2025}]{EMUMainSurvey} 
Hopkins, A.~M., et al., 2025, PASA, 42, e071, \url{https://doi.org/10.1017/pasa.2025.10042}%
\bibitem[\protect\citeauthoryear{Li et al.}{2024}]{LLaVA-OneVision} 
Li, B., et al., \textit{LLaVA-OneVision: Easy Visual Task Transfer}, 2024, \url{https://doi.org/10.48550/arXiv.2408.03326}
\bibitem[\protect\citeauthoryear{Liu et al.}{2023}]{LLaVA} 
Liu, H., et al., \textit{Visual instruction tuning}, 2023, \url{https://doi.org/10.48550/arXiv.2304.08485}%
\bibitem[\protect\citeauthoryear{Norris et al.}{2021}]{Norris2021}
Norris, R.~P., et al., 2021, PASA, 38, e046, \url{https://doi.org/10.1017/pasa.2021.42}%
%\bibitem[\protect\citeauthoryear{Parker et al.}{2024}]{AstroCLIP}
%Parker, L., et al., 2024, MNRAS, 531, 4990, \url{https://doi.org/10.1093/mnras/stae1450}%
\bibitem[\protect\citeauthoryear{Radford et al.}{2021}]{CLIP} 
Radford, A., et al., In International Conference on Machine Learning, 
2021, PMLR, 8748, \url{https://doi.org/10.48550/arXiv.2103.00020}%
\bibitem[\protect\citeauthoryear{Riggi et al.}{2025}]{Riggi2025} 
S. Riggi et al., 2025, PASA, in press, \url{https://doi.org/10.48550/arXiv.2503.23859}%
\bibitem[\protect\citeauthoryear{Wang et al.}{2024}]{Qwen2VL} 
Wang, P., et al., \textit{Qwen2-VL: Enhancing Vision-Language Model's Perception of the World at Any Resolution}, 2024, \url{https://doi.org/10.48550/arXiv.2409.12191}%
\bibitem[\protect\citeauthoryear{Zhou et al.}{2024}]{TinyLLaVA} 
Zhou, B., et al., \textit{TinyLLaVA: A Framework of Small-scale Large Multimodal Models}, \url{https://doi.org/10.48550/arXiv.2402.14289}

\end{thebibliography}
\end{document}